%% file: arxiv.tex
\documentclass{article}

\usepackage{iclr2027_conference,times}
\usepackage[T1]{fontenc}

\input{math_commands.tex}

\usepackage{hyperref}
\usepackage{url}
\usepackage{graphicx}
\usepackage{booktabs}
\usepackage{sidecap}
\usepackage[singlelinecheck=false]{caption}
\usepackage{amsmath}
\usepackage{amssymb}
\usepackage{xcolor}
\IfFileExists{ulem.sty}{\usepackage[normalem]{ulem}}{}   %
\usepackage{multirow}
\usepackage{algorithm}
\usepackage[noend]{algpseudocode}
\usepackage{wrapfig}  %
\algrenewcommand\algorithmicrequire{\textbf{Input:}}
\algrenewcommand\algorithmicensure{\textbf{Returns:}}

\graphicspath{{figures/}}


  \newcommand{\Bm}[1]{\textbf{#1}}

  \newcommand{\keighta}{\space($k{=}8$)}
  \newcommand{\Ba}[1]{#1}\newcommand{\Bam}[1]{$#1$}

  \newcommand{\Bp}[1]{\textbf{#1}}\newcommand{\Bpm}[1]{$\mathbf{#1}$}
  \newcommand{\rowsBG}{4}\newcommand{\rowsSLME}{3}\newcommand{\rowsMH}{3}\newcommand{\rowsBULK}{4}
  \newcommand{\keight}{}   %

\title{Multi-property Feynman-Kac Steering of Crystal Diffusion Models} %
\author{\hfil Oded Mousai$^{1}$ \quad Sapir Esther Yiflach$^{1}$ \quad Barak Itzchak Davidovitch$^{1}$ \\
\hfil\bfseries Yuval Mualem$^{2,3}$ \quad Sarit Kraus$^{1}$ \quad Hannah-Noa Barad$^{2,3}$ \quad Karsten Kreis$^{4}$ \quad Gal Chechik$^{1,4}$ \\[3pt]
\hfil{\normalfont\small $^{1}$Department of Computer Science, Bar-Ilan University} \\
\hfil{\normalfont\small $^{2}$Department of Chemistry, Institute of Nanotechnology \& Advanced Materials, Bar-Ilan University} \\
\hfil{\normalfont\small $^{3}$Israel National Institute for Energy Storage (INIES), Bar-Ilan University} \\
\hfil{\normalfont\small $^{4}$NVIDIA}}

\newcommand{\ourmethod}{FK-ReMask}
\newcommand{\ours}[1]{\textcolor{blue}{#1}}   %

\usepackage{xcolor}
\definecolor{codegreen}{rgb}{0,0.6,0}
\definecolor{codegray}{rgb}{0.5,0.5,0.5}
\definecolor{codepurple}{rgb}{0.58,0,0.82}
\definecolor{backcolour}{rgb}{0.95,0.95,0.92}
\colorlet{reddish}{red!70}    
\colorlet{blueish}{blue!50}    
\colorlet{greenish}{green!60!black}

\newcommand\ignore[1]{{}}

\iclrfinalcopy %

\begin{document}

\maketitle

\lhead{Preprint.}

\begin{abstract} 
Pretrained diffusion models provide powerful priors for generating new materials, including inorganic crystals, but generating useful materials requires satisfying specific chemical and physical properties, from mechanical stiffness to bandgaps. Existing approaches either retrain models for new target properties  or steer only the atom coordinates and lattice during generation, leaving composition unchanged despite its fundamental role in determining physical properties.  We present
\textbf{\ourmethod{}}, an inference-time, gradient-free method that steers both the continuous and discrete channels of a pretrained crystal diffusion model. During generation, \ourmethod{} maintains multiple denoising trajectories and replicates promising candidates using Feynman–Kac particle selection. However, selection alone is insufficient: in the atom-type channel, committed elements cannot be revised, so cloning propagates existing compositions and limits the population's ability to explore new ones. We address this with an atom re-masking mechanism, which reopens committed sites during the generation process and allows alternative compositions to emerge. Across four electronic, photovoltaic, and mechanical objectives, our method improves the yield of SUN (stable, unique, novel) and property-on-target materials. We further demonstrate joint steering of three photovoltaic properties, showing that multiple physical objectives can be specified entirely at inference time without retraining.

\vspace{0.4em}
\centerline{Project page: \url{https://odedmous.github.io/fkremask/}}
\end{abstract}

\section{Introduction}
\label{sec:intro}
Designing new crystal materials with desired properties holds huge promise to advance diverse technologies, from photovoltaic cells, through carbon capture and catalysts, to efficient batteries.  
Diffusion models %
provide a natural framework for crystal generation \citep{zeni2025mattergen}, 
but they typically learn to model the distribution of known materials they were trained on, rather than satisfy desired property requirements.
For example, a photovoltaic absorber may require a suitable bandgap together with favorable optical and electronic properties, while remaining stable and novel. Designing methods that can generate such materials largely remains an open problem.
 
\begin{figure}[!t]
    \centering
    \includegraphics[width=0.99\linewidth]{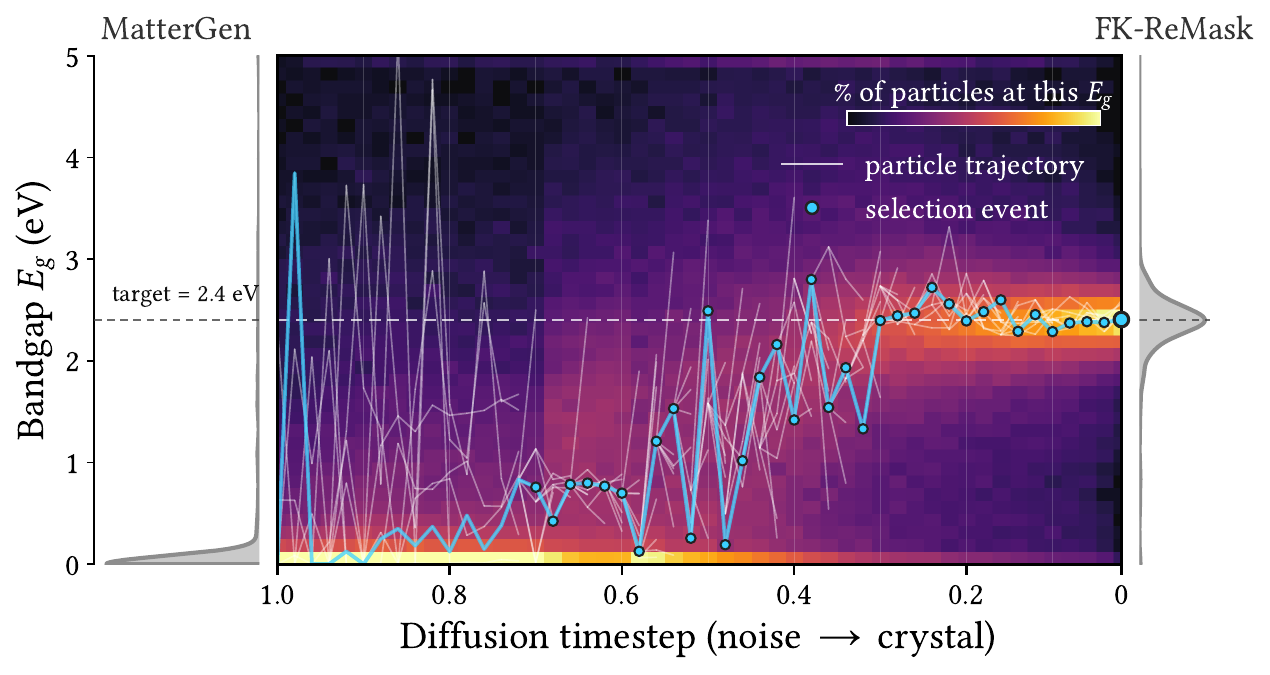}
    \caption{
    \textbf{A snapshot of guiding material generation towards a desired bandgap, with FK sampling.} %
    The figure traces the distribution of (predicted) bandgap as it changes through the diffusion denoising process, from random noise $t=1$ to finished crystals
    $t = 0$. Light blue line traces one particle, with selection events marked with circles; 
    thin white lines trace other particle lineages. The heatmap gives the distribution of the $4096$ live particles at
    each timestep.%
    ; the dashed line marks the target 
     $2.4$\,eV . Once FK selection begins ($t=0.7$) the population shifts towards the target line. %
     Side panels show the BG distribution for 
    unsteered MatterGen (left) and \ourmethod{} (right). }
    \label{fig:method:fk}
\end{figure}

Current diffusion generative approaches target desired properties in two main ways. First, during training, using classifier-free guidance (CFG), input conditioning \citep{zeni2025mattergen}, or reinforcement-learning fine-tuning \citep{chen2025matinvent, guo2026orthogonal}. Second, at inference time, through training-free guidance (steering) \citep{delambilly2026guidance}. Current training-based approaches modify the model for target seen in training, and quite disappointingly, tend to generalize poorly to combinations of multiple properties.%
 Training-free guidance avoids retraining cost by steering with gradients at generation time, but they are limited to differentiable objective. They act only on continuous channels (atom positions and lattice parameters) because a score-based gradient has no straightforward counterpart in the atom-type discrete channel. These approaches therefore do not guide the material composition (atom element type)  %
despite being a fundamental determinant of material properties.

Here we describe  \textbf{FK-ReMask}, an inference-time, gradient-free framework for multi-property control of pretrained crystal diffusion models. It has two important components: FK selection, and atom re-masking.
For the first, FK-ReMask  uses reward-based Feynman–Kac particle selection over a frozen pretrained generator. When generating a crystal, we maintain a small population of  denoising trajectories for parallel exploration, periodically scoring their intermediate structures using external property predictors, and cloning higher-reward trajectories. Since rewards are computed on the full material state, this selection jointly steers the continuous and discrete channels. Multiple properties can be targeted by combining their reward terms, without updating the generator or requiring additional property-labeled data.

Selection alone, however, creates a challenge in the discrete atom channel. The pretrained generator fixes atom types once they are committed, so cloning high-reward particles progressively concentrates the population around previously selected compositions. As a second key component of our method, we therefore use an atom re-masking mechanism, which probabilistically reopens committed sites, allowing selected lineages to explore alternative chemical formulas during the generation process.
Figure~\ref{fig:method:fk} shows an example of bandgap steering with \ourmethod{}:
once selection begins, the population shifts toward the target value.

We test our method on four single-property objectives and a joint three-objective photovoltaic
absorber specification. The latter combines complementary requirements on optical performance,
carrier transport, and electronic structure, demonstrating a realistic setting for target multiple material properties simultaneously.

\paragraph{Our contributions are:} (1) FK-ReMask is the first training-free steering method for crystal diffusion models that jointly guides both the  continuous channel of the material (coordinates and lattice) and its discrete (element composition) channel. 
(2) We introduce a novel use of atom re-masking to mitigate compositional collapse induced by particle cloning, enabling continued exploration of alternative compositions during the generation process.
(3) Across single- and multi-property targeting tasks, FK-ReMask increases the fraction of generated crystals that are simultaneously stable, unique, novel and on the target property.

\newcommand{\unverified}{\textcolor{red}{\textbf{[verify]}}}

\section{Related work}
\label{sec:related}

\noindent\textbf{Particle-based diffusion steering.}
These methods steer generation by selecting among a population of trajectories. Interacting-particle methods provide the foundation for Feynman–Kac and sequential Monte Carlo approaches \citep{delmoral2004fk}, which have been adapted to diffusion models for conditional sampling and inverse problems \citep{trippe2023smcdiff, cardoso2024mcgdiff, wu2023tds}. More recent work uses particle populations explicitly for reward-based steering. \citep{singhal2025fk} formulate Feynman–Kac steering with trajectory-level potentials, and \citep{didi2026proteina} apply it to atomistic binder design. \citet{hasan2026dfkc} use SMC steering in masked discrete diffusion in text generation, where they report population collapse under particle cloning and address it by suspending resampling during the final stage of generation. This is in contrast to our approach which purpose to uses re-masking mechanism to remedy this diversity collapse.

\noindent\textbf{Property targeting in crystal diffusion models.} Most existing approaches reach property targets by modifying or fine-tuning the generator, which is costly in time and property-labeled data, and limits flexibility when targeting new properties or combinations of them. MatterGen\citep{zeni2025mattergen} fine-tunes adapter modules and
conditions through classifier-free guidance, while exploring up to two joint property optimization. MatInvent  \citep{chen2025matinvent} fine-tunes the same backbone with reinforcement learning, reaching up to three simultaneous objectives. As with other training-based methods, %
changing the target value
requires a new fine-tuning run. \citep{guo2026orthogonal} avoid jointly labeled multi-property data by fine-tuning one flow
model per property, then composing the resulting
correction fields at sampling time. New
combinations then require no retraining, but each new property still requires its own
fine-tuning run.

Recent work has also explored inference-time targeting. \citet{delambilly2026guidance} steer MatterGen with Universal Guidance, a form of training-free guidance that requires a differentiable objective. Their updates act on atom positions and lattice parameters and leave the discrete atom channel unguided, so they can reshape geometry but not composition. ALM \citep{edamadaka2026alm} couples a language model to MatterGen through a learned conditioning bridge that injects the text condition into the diffusion process, and introduces T2C-FK, an $N$-particle SMC sampler that resamples trajectories according to rewards evaluated on clean estimates of partially denoised structures. However, they only targrt one physical property, formation energy, while keeping it's composition fixed.

\noindent\textbf{Discrete diffusion and revision operators.}
D3PM \citep{austin2021d3pm} introduces the absorbing corruption used by the discrete atom channel: a site remains masked until it commits to an element, after which its identity cannot be changed. \citet{wang2025remdm} observe that masked discrete diffusion therefore cannot revise a
generated token ``even when it introduces an error'', and derive ReMDM, a principled
re-masking backward process whose operator we adopt. \emph{Remask, Don't Replace} \citep{yao2026remask}, \emph{Revise, Don't Freeze}
\citep{yu2026revise} and ProSeCo \citep{schiff2026proseco} similarly enable previously
generated tokens to be revised in diffusion language models. However, these methods use re-masking to enable revision within a single generation trajectory, while we
use this mechanism to preserve compositional diversity within a group of trajectories.

\section{Preliminaries}
\label{sec:method:setup}

\textbf{Representating crystal materials.} A crystal is represented as a $M = (A, X, L)$, where $A = (a_1, \dots, a_N)$ is the list of atom types at the $N$ sites,
$X$ contains fractional coordinates and $L$ the lattice.

\textbf{Score-based crystal diffusion models.}  It uses a score network $s_\theta$ to jointly denoise the continuous variables $X$ and $L$ with VE-SDE and VP-SDE, respectively, and the discrete atom types $A$ with an absorbing D3PM process. \citep{austin2021d3pm} . This network takes as input crystal $M_t$  and returns continuous scores for $X$ and $L$, together with
logits over clean atom types at each site. A softmax over the non-mask vocabulary, applied
independently at each site, gives $\hat{p}_\theta(a \mid M_t)$, the predicted probability that a
site's clean element is $a$. We keep $s_\theta$ frozen and never differentiate through it.

\section{Our method: \ourmethod{}}
\label{sec:method}

\begin{figure}[t]
    \centering
    \includegraphics[width=0.99\linewidth]{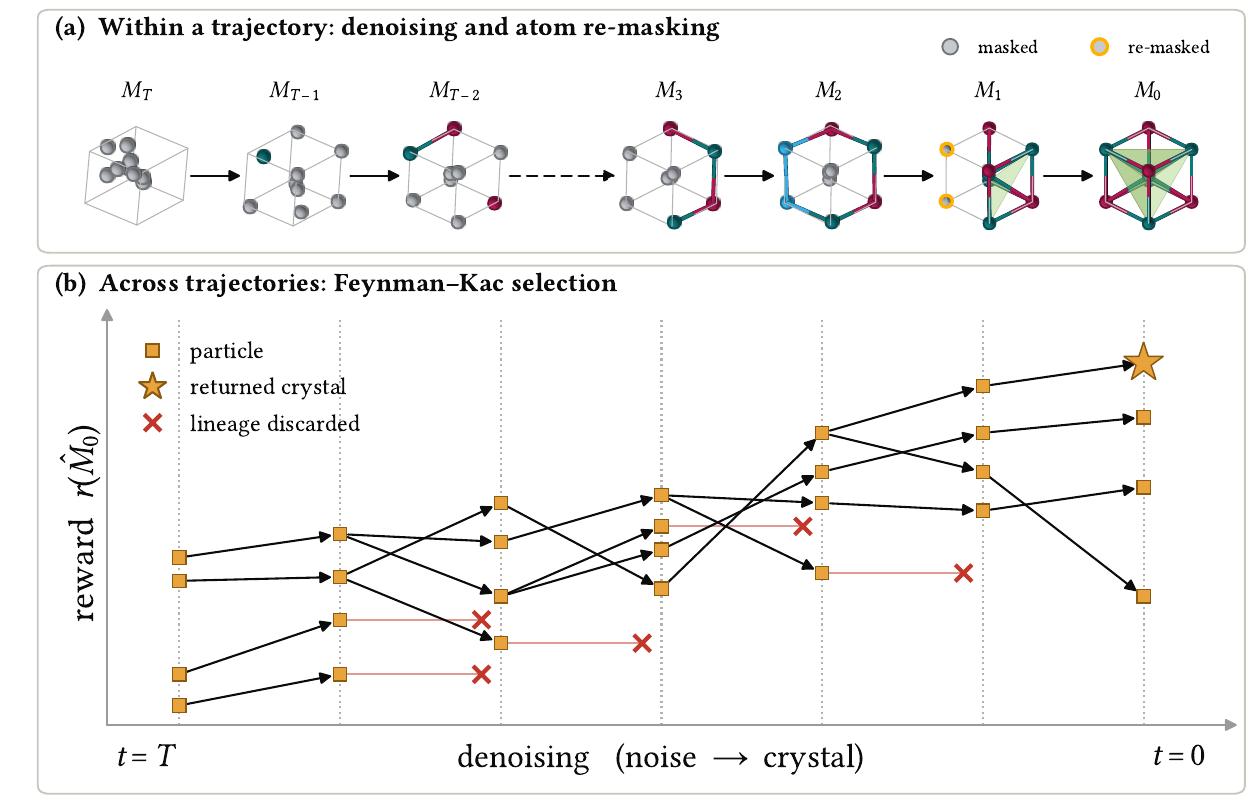}
    \caption{ \textbf{Overview of \ourmethod{}}, combining
    \textbf{(a)} iterative denoising with atom re-masking within single trajectories, and
    \textbf{(b)} reward-guided Feynman–Kac particle selection to prune and clone
    candidates across trajectories.
    }
    \label{fig:method:overview}
\end{figure}

FK-ReMask combines two components for inference-time control of a frozen crystal diffusion model. \textbf{(1) Feynman--Kac Selection} reallocates particles toward higher-reward denoising paths. \textbf{(2) Atom Re-Masking Mechanism} reopens committed atom types, allowing the model to explore compositional alternatives during the generation. Together, they steer both continuous and discrete material channels. Figure~\ref{fig:method:overview} illustrates the  method and Algorithm~\ref{alg:method} gives the full sampler. 

\subsection{Feynman--Kac selection}
\label{sec:method:fk}
\paragraph{Particle selection.}
We formulate inference-time steering as a Feynman--Kac interacting particle system
\citep{delmoral2004fk}. When generating a crystal, we maintain a population of $k$ particles which are parallel evolving under the same frozen reverse dynamics. Steering toward predefined property targets occurs through periodic selection. Every $E$ reverse steps
within a selection window, each particle's noisy state $M_t$ is mapped to an estimated clean crystal
$\hat{M}_0 = (\hat{A}_0, \hat{X}_0, \hat{L}_0)$. \textsc{Tweedie} gives the positions and lattice as
the posterior mean under the model, $(\hat{X}_0, \hat{L}_0) = \mathbb{E}\big[(X_0, L_0) \mid M_t\big]$,
computed from the network's score \citep{efron2011tweedie} (Appendix~\ref{app:tweedie});
\textsc{ImputeMasked} gives the atom types, filling each masked site with
$\argmax_a \hat{p}_\theta(a \mid M_t)$, its most likely clean element, and leaving sites that already
hold an element unchanged. The $k$ slots are then redrawn with probability
proportional to the potential $G^{(i)}_t = \exp\big(r(\hat{M}^{(i)}_0)/\tau\big)$, where $r$
scores a candidate crystal by how close its predicted properties are to the targets, and $\tau$
sets how sharply the group concentrates. The reward is defined below.

This process repeatedly allocates more particles to
promising regions of the generation process while retaining multiple trajectories for exploration.  The reward never sees a noisy state, which
is what lets it be any off-the-shelf predictor trained on finished crystals. Selection occurs only every $E$ steps rather than at every step, 
which lets a selected lineage evolve between these events. At $t = 0$ all the particles are rewarded using their clean finished crystals, and the one with the highest reward is returned.

\begin{algorithm}[t]
\small
\caption{\textbf{\ourmethod{}}. Guided generation of a single crystal. Blue
text marks our changes to standard FK. Every particle update
applies to all $i \in [k]$.}
\label{alg:method}
\begin{algorithmic}[1]
\Require frozen score network $s_\theta$, reward $r$, particles $i=1,...,k$, atom count $N$,
reverse steps $T$, selection interval $E$, temperature $\tau$, re-mask rate $\sigma$, re-masking
$T_{\mathrm{rm}}$ and selection  $T_{\mathrm{sel}}$ windows.
\vspace{0.1pt}

\State $M^{i}_{T} \sim \mathrm{prior}(\cdot \mid N)$
  \Comment{$k$ particles, $N$ atoms each}
\For{$t = T, \dots, 1$}
  \State $X^{i}_{t-1}, L^{i}_{t-1} \gets \textsc{ContinuousDenoise}\big(s_\theta, M^{i}_{t}\big)$
  \State \ours{$A^{i}_{t-1} \gets \textsc{DiscreteDenoise}\big(s_\theta,
    T_{\mathrm{rm}}, M^{i}_{t}\big)$}
    \Comment{\ours{Eq.~\ref{eq:remask}}}
  \State $M^{i}_{t-1} = (A^{i}_{t-1}, X^{i}_{t-1}, L^{i}_{t-1})$
  \If{$t \in T_{\mathrm{sel}}$ \textbf{and} $t \bmod E = 0$}
    \State $\hat{X}^{i}_{0}, \hat{L}^{i}_{0} \gets \textsc{Tweedie}\big(s_\theta, M^{i}_{t-1}\big)$
    \State \ours{$\hat{A}^{i}_{0} \gets \textsc{ImputeMasked}\big(s_\theta, M^{i}_{t-1}\big)$}
      
    \State $\hat{M}^{i}_{0} = (\hat{A}^{i}_{0}, \hat{X}^{i}_{0}, \hat{L}^{i}_{0})$
     \Comment{clean crystal estimation}
    \State $w \gets \mathrm{softmax}\big(r(\hat{M}^{\,\cdot}_{0})/\tau\big)$
      \Comment{over the group}
    \State $\mathrm{pa}(i) \sim \mathrm{Cat}(w)$
      \Comment{$k$ draws with replacement}
    \State $M^{i}_{t-1} \gets M^{\mathrm{pa}(i)}_{t-1}$
  \EndIf
\EndFor
\State \Return $M_{\mathrm{final}} = \arg\max_{i} \, r(M^{i}_{0})$
  \Comment{on finished crystals}
\end{algorithmic}
\end{algorithm}

\label{sec:reward}
\label{sec:method:reward}

\paragraph{Single-property reward.} For a target property, let $v_{\mathrm{pred}}(M)$ denote the predicted value and $v_{\mathrm{target}}$ the desired target. Assuming a Laplace prior distribution around the target property,
$p \propto \exp\big(-\lvert v_{\mathrm{pred}}(M) - v_{\mathrm{target}}\rvert / s\big)$,
our reward is the log likelihood
$r = \log p =-\lvert v_{\mathrm{pred}}(M) - v_{\mathrm{target}}\rvert / s$.  The rescaling factor $s$ is computed once from samples generated by the unsteered model and then fixed throughout steering,
$s = \mathrm{sd}\big(\lvert v - v_{\mathrm{target}}\rvert\big)$,  putting reward
terms for different properties on a comparable scale.
A property may be defined only on part of the space; for example, a property derived from a
band edge is meaningless for a metal. Writing $\mathcal{D}$ for that domain, we set
$p(M)=\varepsilon>0$ outside it, which avoids $\log 0 = -\infty$ and keeps rewards
non-degenerate across invalid candidates.

\paragraph{Multi-property reward.} Rather than combining objectives
heuristically, we define the distribution we intend to sample. Treating each property as a
soft constraint on the pretrained prior $p_\theta$ gives a product of experts,
\begin{align}
  p(M) \;&\propto\; p_\theta(M) \prod_{j=1}^{P} p_j(M), \label{eq:poe}\\[4pt]
  p_j(M) \;&\propto\;
  \begin{cases}
    \exp\!\big(-\lvert v^j_{\mathrm{pred}}(M) - v^j_{\mathrm{target}}\rvert / s^j\big),
      & M \in \mathcal{D}_j,\\[4pt]
    \varepsilon, & \text{otherwise},
  \end{cases} \label{eq:expert}
\end{align}
so that the log-likelihood reward is
\begin{equation}
  r(M) \;=\; \sum_{j=1}^{P} r_j(M), \qquad r_j(M) = \log p_j(M).
  \label{eq:reward-multi}
\end{equation}

Table~\ref{tab:app:rewards} gives each domain $\mathcal{D}_j$ and the value of $\varepsilon$.

Note that Eq.~\eqref{eq:poe} treats each property as a separate filter. This is convenient but
not accurate because some of the properties we target are physically coupled. We find this
sufficient in practice, and leave correlation-aware experts and non-linear scalarizations to
future work.

\subsection{Atom re-masking}
\label{sec:method:remask}
Selection only samples from crystals that the  population already holds, limiting exploration. The continuous channels in our model keep exploring: every
reverse step draws fresh Gaussian noise per particle, so positions and lattices go on moving. The
atom-type channel does not. It is absorbing, so once a site commits to an element no later step can
change it, and selection over composition reduces to choosing among formulas committed earlier in
the trajectory. That is a real limit, since the properties we target depend on composition as much
as on geometry.

Resampling every committed site would restore that freedom, but it would discard the composition
built so far and leave a masked fraction the frozen model was never trained on. We reopen only part
of it instead, controlled by a re-masking rate $\sigma$, and compensate so that the schedule is preserved.

Formally, let $\gamma_t$ be the fraction of sites committed at time $t$, so that $\gamma_{t-1} > \gamma_t$ at the next, less noisy step, and write $\delta_x$ for the point mass on $x$.  Under the stock
absorbing kernel a committed site is frozen and a masked site unmasks with probability
$(\gamma_{t-1}-\gamma_t)/(1-\gamma_t)$. We replace it with an \textbf{atom re-masking} kernel,
applied independently to each of the $N$ sites:
\begin{equation}
q_{\sigma}\big(a_{t-1} \mid a_{t}\big) =
\begin{cases}
(1-\sigma_t)\,\delta_{a_t} \;+\; \sigma_t\,\delta_{[\mathrm{MASK}]},
  & a_{t} \neq [\mathrm{MASK}], \\[6pt]
\dfrac{\gamma_{t-1} - (1-\sigma_t)\,\gamma_t}{1-\gamma_t}\; \hat{p}_\theta
  \;+\; \dfrac{1-\gamma_{t-1}-\sigma_t\,\gamma_t}{1-\gamma_t}\; \delta_{[\mathrm{MASK}]},
  & a_{t} = [\mathrm{MASK}],
\end{cases}
\label{eq:remask}
\end{equation}
with the rate gated to the window and capped so the mask coefficient stays non-negative,
\begin{equation}
\sigma_t \;=\; \min\!\Big\{\, \sigma \cdot \mathbf{1}\big[\,t \in T_{\mathrm{rm}}\,\big],
\;\; \tfrac{1-\gamma_{t-1}}{\gamma_t} \,\Big\},
\qquad\text{so } \sigma_t = 0 \text{ outside } T_{\mathrm{rm}}.
\label{eq:sigma}
\end{equation}
The unmasking rate in the second branch is the stock rate plus $\sigma_t\gamma_t/(1-\gamma_t)$,
exactly the mass returned to the mask state spread over the $1-\gamma_t$ masked sites, so the
committed fraction is unchanged in expectation:
$\gamma_t(1-\sigma_t) + (\gamma_{t-1} - (1-\sigma_t)\gamma_t) = \gamma_{t-1}$. Because
unmasking is conditioned on the surrounding structure, variation enters without leaving the model's
own distribution.

\begin{figure}[h]
    \centering
    \includegraphics[width=\linewidth]{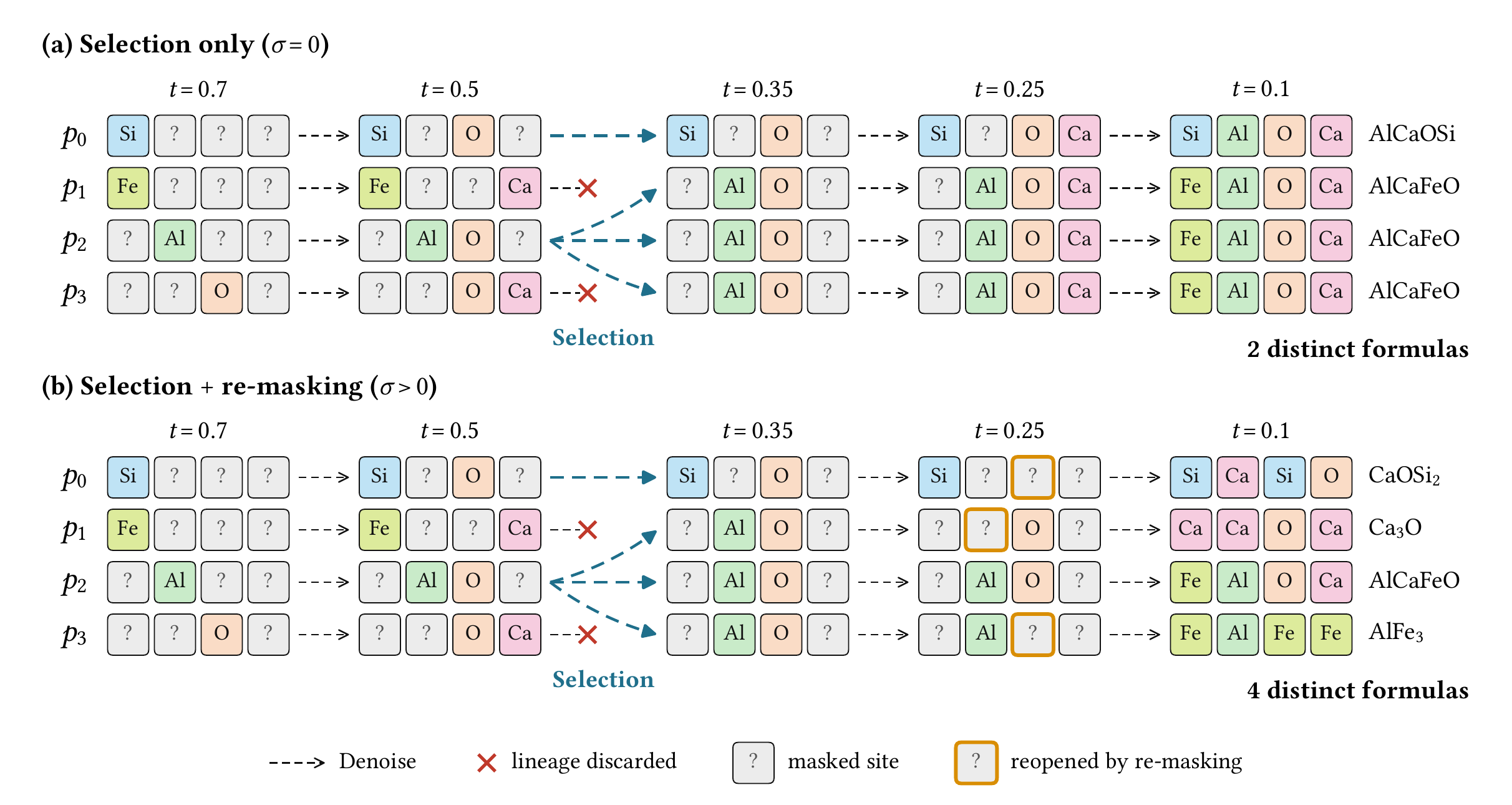}
    \caption{\textbf{Illustration of atom re-masking under selection.}  Four particles go through denoising, selection, and re-masking.%
    Cells denote the element composition at atom sites. 
    \textbf{(a)} Without re-masking, $\sigma = 0$, the selection step at $t = 0.35$ allocate  winning particles to more slots. This reduces the diversity of the final generated compositions. \textbf{(b)} With re-masking, $\sigma > 0$, committed sites reopen
    inside the re-mask ($t=0.25$) window yielding  four distinct formulas. %
    }
    \label{fig:remask:schematic}
\end{figure}

We adopt this operator from ReMDM \citep{wang2025remdm} but repurpose it: they re-mask to correct
decoding errors within one trajectory, whereas we use it against the compositional collapse that cloning induces across trajectories (Figure~\ref{fig:remask:schematic}).

\section{Experimental setup and details}
\label{sec:setup}
\label{sec:setup:eval}

\paragraph{Experimental Design.} We evaluate FK-ReMask against  baseline methods across four material properties, first in single-objective settings (\S\ref{sec:results:single}) and then by simultaneously targeting three photovoltaic properties (\S\ref{sec:multiobj}). We use MatterGen as our pretrained unconditional crystal diffusion model. 

\paragraph{Material Properties.} We consider three photovoltaic properties and one mechanical property.  
\textbf{(1)~Bandgap (BG)} determines the spectral range available for absorption and constrains the achievable voltage, making it useful for identifying materials with suitable electronic structure for solar-energy conversion.
\textbf{(2)~Solar-cell efficiency (SLME)} 
accounts for the absorption spectrum and the distinction between fundamental and direct-allowed gaps 
\citep{yu2012slme}. 
 \textbf{(3) Hole effective mass} captures carrier transport, as large effective masses can hinder carrier extraction. Together, BG, SLME, and hole effective mass capture complementary requirements for a practical solar absorber. \textbf{(4) Bulk modulus} characterizes resistance to volumetric compression, providing a target distinct from the electronic and optical properties. 

\begin{table}[h]
\caption{Properties, target values, and success criteria.}
    \label{tab:setup:properties}
    \centering
    \setlength{\tabcolsep}{4pt}
    \begin{tabular}{@{}lll@{}}
    \toprule
    \textbf{Material property} & \textbf{$v_{\mathrm{target}}$} & \textbf{On-target criteria} \\
    \midrule
    Bandgap ($E_g$) &
      2.4\,eV (\S\ref{sec:results:single}), 1.3\,eV (\S\ref{sec:multiobj:solar}) &
      $\lvert E_g - v_{\mathrm{target}} \rvert \leq 0.1$\,eV \\
    Solar-Cell Efficiency (SLME) &
      maximise ($33.7\,\%$ is the ceiling) &
      $E_g \geq 0.5$\,eV $\wedge$ $\mathrm{SLME} \geq 25\,\%$ \\
    Bulk Modulus (K) &
      150\,GPa &
      $\lvert K - 150 \rvert \leq 15$\,GPa \\
    Hole Effective Mass ($m^{*}_h$) &
      minimise ($0$ is the floor) &
      $E_g \geq 0.5$\,eV $\wedge$ $0 \leq m^{*}_h \leq 0.3$ \\
      \midrule
    \bottomrule
    \end{tabular}
\end{table}

\paragraph{Metrics.} We report the percentages of generated structures that are valid, unique, novel, stable ($E_{\mathrm{hull}} \leq 0.1$\, eV/atom against the MP2020 convex hull) and SUN (structures that are simultaneously stable, unique, and novel). We further evaluate how well the generation met target property, using three : (1) on-target - the fraction of materials that satisfy the specified property target; (2) stable \& on-target; and (3) SUN \& target. Target values and property-specific criteria are given in Table~\ref{tab:setup:properties}.

\paragraph{Evaluation Protocol.} We followed the MatterGen evaluation pipeline. For each method in each experiment, we generate $n=512$ structures, relax them, and calculate the evaluation metrics on the relaxed structures. Structures that the evaluator cannot process are retained as failures rather than removed from the evaluation set. The on-target metric is computed using off-the-shelf pretrained property predictors, see Appendix~\ref{app:predictors} for details. 

\paragraph{Baselines.} We compare \ourmethod{} with three baselines: \textbf{(1) MatterGen unsteered} samples from the frozen pretrained model without
conditioning. \textbf{(2) MatterGen with CFG} using the released classifier-free-guidance adapter, fine-tuned on property-labelled data,  available only for for bandgap and bulk modulus.
\textbf{(3) Best-of-N}. For each of $n$ outputs, we generate $k$ independent samples from the unsteered MatterGen model and select the best after relaxation under \ourmethod{}'s own reward, matching FK's $kn$ trajectories. 

\paragraph{Hyperparameters.} We use $k = 8$ particles per generated crystal. Particle in 
a group are initialized from the prior with a shared atom count $N$. Selection occurs every $E = 20$ reverse steps at
temperature $\tau = 0.2$ within a window $T_{\mathrm{sel}} = [0.7, 0]$, giving 35 events on a
uniform grid. Atom re-masking is applied at every reverse step within a window
$T_{\mathrm{rm}} = [0.60, 0.05]$, each committed site returning to the mask state with
rate $\sigma = 0.1$.

\section{Results}
\label{sec:results}
We first evaluate steering generation using a single target property, repeated with 4 properties, then steer towards multiple properties jointly. 
\subsection{Optimizing a single requested property}
\label{sec:results:single}

\begin{figure}[ht]
    \centering
    \includegraphics[width=0.85\linewidth]{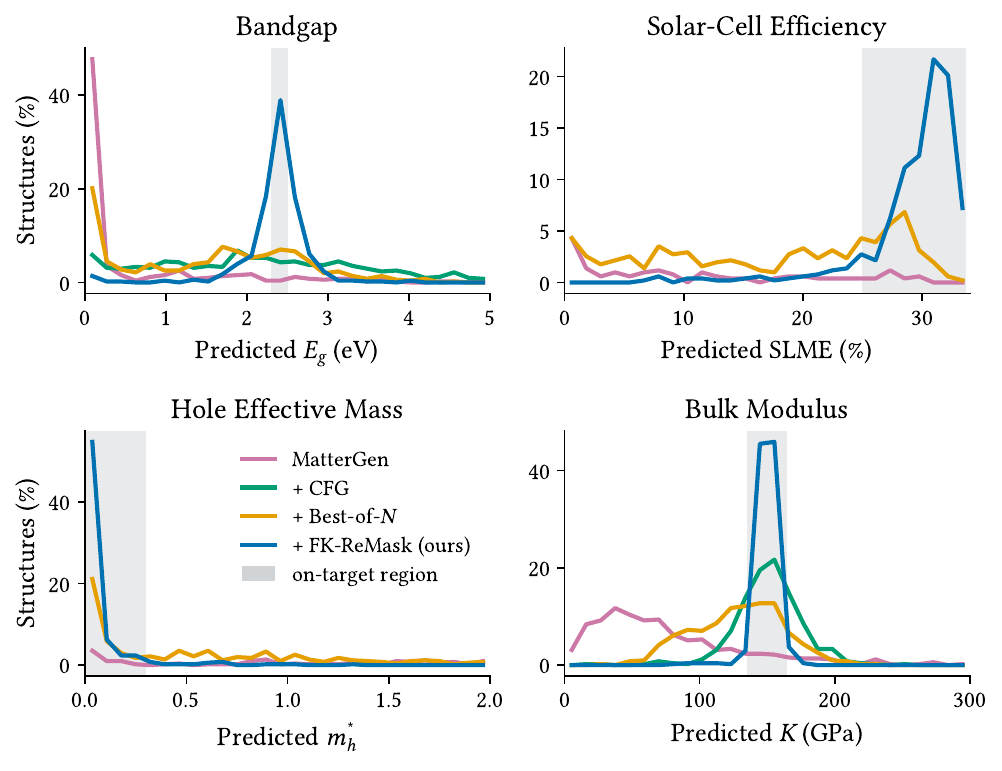}
    \caption{\textbf{Distribution of target  properties}. Curves show histogram of properties obtained be generating  crystals using various methods. In the bottom two panels, \textit{Solar-cell efficiency} and \textit{Hole Effective Mass} metals were removed from the distribution since these properties are irrelevant for metals, and only gapped structures are plot  ($E_g \geq 0.5$\,eV). The percentages remain over all n=512. Their curves therefore sum to each method's gapped fraction rather than to $100\%$.
    } 
    \label{fig:results:densities}
\end{figure}

Table~\ref{tab:results:all} compares \ourmethod{} to the baselines on four target properties. 
Among the inference-time methods, \ourmethod{} gives the highest stable \& on-target yield on all four properties, by
 30.2 to 51.2 percentage points over the compute-matched baseline%
.  %
Figure~\ref{fig:results:densities} shows the 
distributions of predicted values for all methods and target properties.

\begin{table}[h]
\caption{Single-objective results across four properties. 
Bold indicates the best value for each metric within each property.}
    \label{tab:results:all}
    \centering
    \small
    \setlength{\tabcolsep}{2.4pt}
    \begin{tabular}{llrrrrrrrr}
    \toprule
    property & method & valid & unique & novel & stable & SUN & \shortstack{on\\target} & \shortstack{stable \&\\on target} & \shortstack{SUN \&\\on target} \\
    \midrule
    \multirow{\rowsBG}{*}{\shortstack[l]{Bandgap\\($E_g$)}}
 & MatterGen & \Bp{86.3} & 99.6 & 59.2 & 75.0 & 37.5 & 0.4 & $0.4${\scriptsize$\pm0.3$} & $0.2${\scriptsize$\pm0.2$} \\
 & $+$ CFG & \Bp{86.3} & 99.4 & 67.0 & 67.2 & 36.5 & 4.9 & $3.1${\scriptsize$\pm0.8$} & $1.2${\scriptsize$\pm0.5$} \\
 & $+$ Best-of-N\keight & 82.8 & \textbf{100.0} & \Bp{68.4} & 69.7 & 41.4 & 7.6 & $5.5${\scriptsize$\pm1.0$} & $3.9${\scriptsize$\pm0.9$} \\
 & $+$ \textbf{FK-ReMask}\keight & 78.1 & 96.5 & 67.0 & \Bp{82.6} & \Bp{51.2} & \Bp{42.8} & \Bpm{35.7}{\scriptsize$\pm2.1$} & \Bpm{20.3}{\scriptsize$\pm1.8$} \\
    \midrule
    \multirow{\rowsSLME}{*}{\shortstack[l]{Solar-Cell \\ Efficiency (SLME)}}
 & MatterGen & \Bp{86.3} & 99.6 & 59.2 & 75.0 & 37.5 & 2.5 & $1.8${\scriptsize$\pm0.6$} & $1.2${\scriptsize$\pm0.5$} \\
 & $+$ Best-of-N\keight & 85.2 & \Bp{99.8} & 65.6 & 69.1 & 37.5 & 24.0 & $16.2${\scriptsize$\pm1.6$} & $9.6${\scriptsize$\pm1.3$} \\
 & $+$ \textbf{FK-ReMask}\keight & 81.2 & 98.0 & \Bp{67.8} & \Bp{81.2} & \Bp{49.2} & \Bp{82.0} & \Bpm{67.4}{\scriptsize$\pm2.1$} & $\mathbf{42.4}${\scriptsize$\pm2.2$} \\
    \midrule
    \multirow{\rowsMH}{*}{\shortstack[l]{Hole Effective \\Mass ($m^{*}_h$)}}
 & MatterGen & \Bp{86.3} & 99.6 & 59.2 & 75.0 & 37.5 & 5.7 & $3.7${\scriptsize$\pm0.8$} & $2.9${\scriptsize$\pm0.7$} \\
 & $+$ Best-of-N\keight & 84.6 & \Bp{99.8} & \Bp{65.6} & 69.9 & 38.7 & 31.6 & $21.7${\scriptsize$\pm1.8$} & $13.9${\scriptsize$\pm1.5$} \\
 & $+$ \textbf{FK-ReMask}\keight & 84.6 & 95.7 & 62.7 & \Bp{82.4} & \Bp{47.5} & \Bp{65.6} & \Bpm{54.3}{\scriptsize$\pm2.2$} & \Bpm{33.2}{\scriptsize$\pm2.1$} \\
    \midrule
    \multirow{\rowsBULK}{*}{\shortstack[l]{Bulk Modulus\\($K$)}}
 & MatterGen & \Bp{86.3} & \textbf{99.6} & 59.2 & \Bp{75.0} & \Bp{37.5} & 6.4 & $4.1${\scriptsize$\pm0.9$} & $1.8${\scriptsize$\pm0.6$} \\
 & $+$ CFG & 81.1 & \textbf{99.6} & 57.6 & 65.0 & 27.1 & 53.3 & $36.1${\scriptsize$\pm2.1$} & $15.4${\scriptsize$\pm1.6$} \\
 & $+$ Best-of-N\keight & 81.4 & \textbf{99.6} & \Bp{61.7} & 60.5 & 27.5 & 34.4 & $20.5${\scriptsize$\pm1.8$} & $7.8${\scriptsize$\pm1.2$} \\
 & $+$ \textbf{FK-ReMask}\keight & 79.1 & 98.4 & 52.3 & 73.8 & 31.6 & \Bp{95.7} & \Bpm{70.5}{\scriptsize$\pm2.0$} & \Bpm{30.1}{\scriptsize$\pm2.0$} \\
    \bottomrule
    \end{tabular}
\end{table}

\subsection{Multiple properties}
\label{sec:multiobj}

\begin{figure}[t]
\centering
\includegraphics[width=0.99\linewidth]{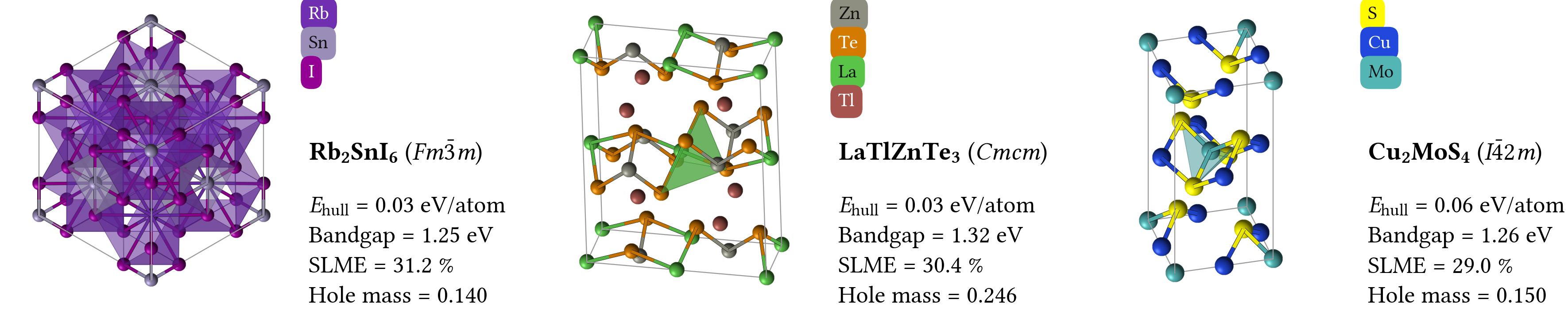}
\caption{\textbf{Three example crystals generated by \ourmethod{}.} %
Property values were predicted using
pretrained ML predictors. Structures are higlighy diverse and meet the desired specification.}
\label{fig:solar:structures}
\end{figure}

We next targeted three properties jointly: bandgap, SLME and hole effective mass, which for
\ourmethod{} amounts to summing their rewards. Table~\ref{tab:multiobj:solar} reports the results.
Both the efficiency and the hole-mass criteria require a bandgap of at least $0.5$\,eV, and
$80.9\%$ of the prior's samples fall short, failing both at once. Those that do clear the floor
land well above the $1.3$\,eV target, so selection has almost nothing to choose from.
Figure~\ref{fig:solar:structures} shows three of \ourmethod{}'s on-target crystals. 

\label{sec:multiobj:solar}

\begin{table}[h]
\caption{Multi-properties results. Jointly targeting three photovoltaic objectives. \emph{all} denotes the conjunction of the three on-target conditions.}
\label{tab:multiobj:solar}
\centering
\setlength{\tabcolsep}{2.5pt}
\resizebox{\linewidth}{!}{%
\begin{tabular}{lrrrrrrrrrrr}
\toprule
 & & & & & & \multicolumn{4}{c}{on target} & & \\
\cmidrule(lr){7-10}
method & valid & unique & novel & stable & SUN & $E_g$ & SLME & $m^{*}_h$ & all & stable \& all & SUN \& all \\
\midrule
MatterGen & \Bp{86.3} & 99.6 & 59.2 & 75.0 & 37.5 & 1.0 & 2.5 & 5.7 & 0.2 & $0.0${\scriptsize$\pm0.0$} & $0.0${\scriptsize$\pm0.0$} \\
$+$ Best-of-N\keight & 83.8 & \Bp{99.8} & \Bp{66.8} & 66.6 & 36.7 & 7.6 & 21.3 & 26.0 & 1.4 & $1.0${\scriptsize$\pm0.4$} & $0.4${\scriptsize$\pm0.3$} \\
$+$ \textbf{FK-ReMask}\keight & 84.0 & 97.5 & 66.4 & \Bp{77.5} & \Bp{46.1} & \Bp{27.0} & \Bp{68.8} & \Bp{58.6} & \Bp{11.5} & \Bpm{9.2}{\scriptsize$\pm1.3$} & \Bpm{5.3}{\scriptsize$\pm1.0$} \\
\bottomrule
\end{tabular}
}
\end{table}

\subsection{Selection is not enough}
\label{sec:remask:required}
\label{sec:abl:rho}

The results above show that selection with atom re-masking improves targeting, but leave open
whether re-masking is necessary.
To isolate this effect, we compare two otherwise identical bandgap steering experiments targeting
$2.4$\,eV: one with re-masking ($\sigma = 0.1$) and one with selection alone ($\sigma = 0$).
Along each denoising trajectory, we measure two quantities: (1) the number of distinct committed
compositions among the eight particles. This measures the chemical alternatives actually
represented in the particle population, rather than those introduced by imputing unresolved sites.
(2) The mean bandgap reward across the particles, computed from the complete estimated structures,
which tracks the quality of the current search population.

\begin{figure}[h]
\centering
\includegraphics[width=0.85\linewidth]{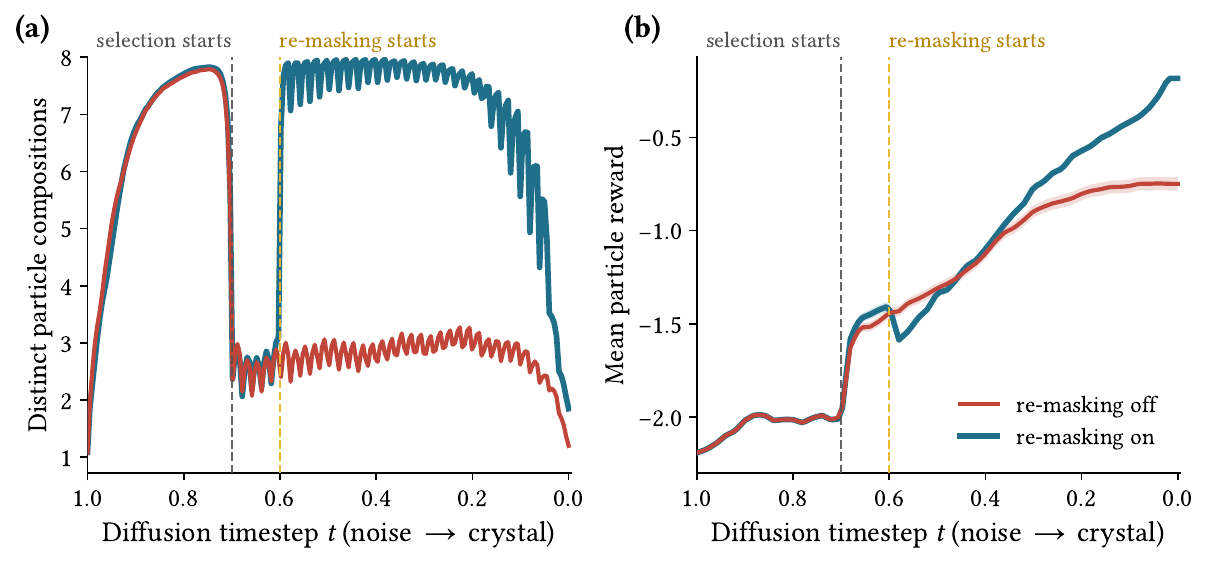}
\caption{\textbf{Bandgap steering
with and without re-masking.} Vertical dashed lines mark where selection and re-masking begin. \textbf{(a)} Selection collapses the
population, but with re-masking on, diversity is restored. \textbf{(b)} Re-masking reaches higher
mean particle reward.}
\label{fig:remask}
\end{figure}

As shown in Figure~\ref{fig:remask}(a), in both setups particles diverge freely prior to selection, reaching near the maximum of distinct compositions. But the initial selection events trigger a severe collapse in compositional diversity. Without re-masking, the population remains collapsed. In contrast, activating re-masking within $t\in[0.6,0.05]$ allows committed sites to be iteratively revised, sustaining substantially greater compositional diversity across the window. This improves generation quality, raising the terminal mean reward  (Figure~\ref{fig:remask}(b)). Table~\ref{tab:abl:rho:multi} shows the same effect on the three-objective solar experiment from \S\ref{sec:multiobj:solar}: without re-masking, all three targeting rates fall, and the SUN\&on-target yield drops from $5.3\%$ to $1.6\%$.

\begin{table}[h!]
\caption{Re-masking on and off for the three-objective arm.}
\label{tab:abl:rho:multi}
\centering
\setlength{\tabcolsep}{3pt}
\setlength{\aboverulesep}{0.2ex}
\setlength{\belowrulesep}{0.2ex}
\resizebox{\linewidth}{!}{%
\begin{tabular}{lrrrrrrrrrrr}
\toprule
 & & & & & & \multicolumn{4}{c}{on target} & & \\
\cmidrule(lr){7-10}
 & valid & unique & novel & stable & SUN & $E_g$ & SLME & $m^{*}_h$ & all & \shortstack{stable \&\\all} & \shortstack{SUN \&\\all} \\
\midrule
$\sigma = 0$ (selection only) & 83.6 & \textbf{99.6} & 63.1 & 73.2 & 38.9 & 13.3 & 33.6 & 29.5 & 2.7 & $2.3${\scriptsize$\pm0.7$} & $1.6${\scriptsize$\pm0.5$} \\
$\sigma = 0.1$ (ours)         & \textbf{84.0} & 97.5 & \textbf{66.4} & \textbf{77.5} & \textbf{46.1} & \textbf{27.0} & \textbf{68.8} & \textbf{58.6} & \textbf{11.5} & $\mathbf{9.2}${\scriptsize$\pm1.3$} & $\mathbf{5.3}${\scriptsize$\pm1.0$} \\
\bottomrule
\end{tabular}
}
\end{table}

\section{Limitations}
\label{sec:limitations}
\ourmethod{} is designed as a fast, training-free method, that guides generation during inference. The method has several limitations. First, when large datasets of materials with desired target properties are available, finetuning models may end up performing better. Second, 
\ourmethod{} repeatedly calls property predictors in the loop, thereby depending on available light-weight but accurate predictors. Also, real chemical validation often requires complex synthesis protocols that may be hard to achieve for novel materials. Incorporating high-fidelity evaluators that are also aware of synthesizability during denoising could also provide more useful generated materials.

\section{Conclusion}
\label{sec:conclusion}

We introduced \ourmethod{}, a training-free,  inference-time method for steering a frozen crystal generator across its continuous and discrete channels towards  desired material properties. \ourmethod{} dramatically increases the fraction of generated crystals that meet the desired specification, for example a $5\times$ gain in SUN and on-target yield on bandgap. Our approach introduces two key components into diffusion-based material generation: FK selection and atom re-masking.

\section*{Acknowledgments}

This research was supported by the Israel Science Foundation (ISF), grant no. 208075.

\clearpage
\bibliography{references}
\bibliographystyle{iclr2027_conference}

\clearpage
\appendix
\section*{Appendix}
\setcounter{table}{0}
\renewcommand{\thetable}{A\arabic{table}}
\setcounter{figure}{0}
\renewcommand{\thefigure}{A\arabic{figure}}

\section{Property predictors and CFG checkpoints}
\label{app:detail}
\label{app:predictors}

Every reward is one forward call to an off-the-shelf pretrained ALIGNN model from the public zoo;
none is trained or fine-tuned by us. Table~\ref{tab:app:predictors} names them. The two
released MatterGen classifier-free-guidance adapters used as baselines are
\texttt{dft\_band\_gap} for the bandgap and \texttt{ml\_bulk\_modulus} for bulk modulus; no
adapter is released for SLME or hole effective mass.

\begin{table}[h]
\caption{The predictor behind each objective of Table~\ref{tab:setup:properties}.}
\label{tab:app:predictors}
\centering\small
\begin{tabular}{@{}ll@{}}
\toprule
property & predictor \\
\midrule
Bandgap ($E_g$)                & \texttt{mp\_gappbe\_alignn} \\
Solar-Cell Efficiency (SLME)    & \texttt{jv\_slme\_alignn} \\
Bulk Modulus (K)                & \texttt{jv\_bulk\_modulus\_kv\_alignn} \\
Hole Effective Mass ($m^{*}_h$) & \texttt{jv\_avg\_hole\_mass\_alignn} \\
\bottomrule
\end{tabular}
\end{table}

\section{Clean-crystal estimate}
\label{app:tweedie}

Selection scores a particle by the reward of its estimated clean crystal
$\hat{M}_0 = (\hat{A}_0, \hat{X}_0, \hat{L}_0)$, never by a noisy state (\S\ref{sec:method:fk}).
The continuous channels of that estimate come from the Tweedie formula
\citep{efron2011tweedie}: for a Gaussian forward process
\begin{equation}
z_t \mid z_0 \;\sim\; \mathcal{N}\big(\alpha_t z_0 + (1-\alpha_t)\mu,\ \nu_t^2 I\big),
\label{eq:tweedie-forward}
\end{equation}
the posterior mean of the clean variable is available in closed form from the score,
\begin{equation}
\hat{z}_0 \;=\; \mathbb{E}\big[z_0 \mid z_t\big]
\;=\; \frac{z_t \;+\; \nu_t^2\, \nabla_{z_t} \log p_t(z_t) \;-\; (1-\alpha_t)\,\mu}{\alpha_t},
\label{eq:tweedie}
\end{equation}
and we evaluate it with the frozen network's score in place of $\nabla_{z_t} \log p_t$. The two
continuous channels are the two settings of $(\alpha_t, \mu)$. Fractional coordinates follow a
variance-exploding SDE, which does not shrink the signal, so $\alpha_t = 1$ and $\mu = 0$ and
Eq.~\ref{eq:tweedie} reduces to its first two terms,
\begin{equation}
\hat{X}_0 \;=\; X_t \;+\; \nu_t^2\, s_\theta(M_t),
\label{eq:tweedie-pos}
\end{equation}
wrapped back into $[0,1)$. The lattice follows a variance-preserving SDE with $\alpha_t < 1$ and a
non-zero limit mean, $\mu$ being the cubic cell whose volume matches the atom count at the
training set's mean atomic density, so both correction terms of Eq.~\ref{eq:tweedie} are active:
\begin{equation}
\hat{L}_0 \;=\; \frac{L_t \;+\; \nu_t^2\, s_\theta(M_t) \;-\; (1-\alpha_t)\,\mu}{\alpha_t}.
\label{eq:tweedie-cell}
\end{equation}
Here $\nu_t$ is the forward process standard deviation at time $t$; it is unrelated to the
re-masking rate $\sigma_t$ of \S\ref{sec:method:remask}.

Atom types are not covered by Eq.~\ref{eq:tweedie}, which is a statement about Gaussian noise.
They are discrete and corrupted by absorbing masking, so a masked site has no posterior mean to
read off a score. \textsc{ImputeMasked} instead takes the most likely clean element at each masked
site under the network's own categorical head, $\argmax_a \hat{p}_\theta(a \mid M_t)$, and leaves
already committed sites alone. The estimate is therefore a mean on the continuous channels and a
mode on the discrete one. Both are point estimates computed in one forward call, which is what
keeps selection cheap: the reward sees a finished crystal, so any predictor trained on finished
crystals applies unchanged.

\section{Reward terms}
\label{app:rewards}

Table~\ref{tab:app:rewards} gives the reward term used for each objective.

\begin{table}[h]
\centering
\small
\begin{tabular}{@{}ll@{}}
\toprule
Property & Reward $r(M)$ \\
\midrule
Bandgap &
  $-\,\bigl|E_g(M)-v_{\mathrm{bg}}\bigr| \,/\, s_{\mathrm{bg}}$ \\[4pt]
Solar-cell efficiency &
  $\begin{cases}
     -\,\bigl|\mathrm{SLME}(M)-v_{\mathrm{slme}}\bigr| \,/\, s_{\mathrm{slme}},
       & E_g(M)\ge 0.5\,\mathrm{eV}\\
     -5, & \text{otherwise}
   \end{cases}$ \\[10pt]
Hole effective mass &
  $\begin{cases}
     -\,\bigl|m^{*}_h(M)-v_{\mathrm{hm}}\bigr| \,/\, s_{\mathrm{hm}},
       & E_g(M)\ge 0.5\,\mathrm{eV}\\
     -5, & \text{otherwise}
   \end{cases}$ \\[10pt]
Bulk modulus &
  $-\,\bigl|K(M)-v_{\mathrm{bulk}}\bigr| \,/\, s_{\mathrm{bulk}}$ \\
\bottomrule
\end{tabular}
\caption{Per-property reward terms. Solar-cell efficiency and hole effective mass are undefined for metals, although the predictor still returns values, so these terms are restricted to structures with a bandgap of at least 0.5 eV. Outside this region, the corresponding reward term is set to $\log\varepsilon=-5$, ensuring that gapless structures cannot outrank structures with a valid bandgap.}
\label{tab:app:rewards}
\end{table}

\section{Two-objective steering: low supply-chain risk permanent magnets}
\label{app:magnet}

A permanent magnet holds a magnetic field with no power supplied to it, which puts one in every
electric-vehicle traction motor and direct-drive wind-turbine generator. The strongest grades, the
ones electrification is driving demand for, reach their magnetization density $\rho_{\mathrm{mag}}$
(net magnetic moment per unit cell volume) with rare-earth elements such as Nd and Dy, whose
production is concentrated in a few countries. The Herfindahl-Hirschman index (HHI) measures that
concentration as the sum of squared national shares of an element's reserves, averaged over a
compound's elements by mass \citep{gaultois2013hhi}. A candidate worth synthesising has to be strong
and dispersed at once, which makes this a steering problem rather than two separate ones.

MatterGen releases exactly one adapter conditioned on two properties jointly,
\texttt{dft\_mag\_density\_hhi\_score} \citep{zeni2025mattergen}, so this task is the only one
where \ourmethod{} can be compared against a released multi-property adapter. We use its
thresholds, $\rho_{\mathrm{mag}} > 0.2\,\mu_B/\text{\AA}^3$, comparable to elemental iron, and
$\mathrm{HHI} < 1500$, the conventional bound for an unconcentrated market. We predict
$\rho_{\mathrm{mag}}$ with CHGNet \citep{deng2023chgnet} (summed site moments divided by cell volume)
and compute HHI from the composition, reproducing MatterGen's own \texttt{hhi\_score} label (the
reserve index, averaged by weight fraction and renormalised over the elements pymatgen tabulates),
so that steering and the trained adapter optimise the same quantity. The CFG adapter is conditioned
on the same two values, with guidance strength $2.0$.

Both requirements here are thresholds rather than values to hit, so the reward takes a different
shape from the point targets of Appendix~\ref{app:rewards}. Each term measures the distance
\emph{outside} its acceptance region and is therefore zero once that requirement is met, which
stops selection from spending its budget on an objective that is already satisfied. 
\begin{equation}
r(M) \;=\; -\,\frac{\max\bigl(0,\; 0.2-\rho_{\mathrm{mag}}(M)\bigr)}{s_{\mathrm{mag}}}
       \;-\; \frac{\max\bigl(0,\; \mathrm{HHI}(M)-1500\bigr)}{s_{\mathrm{HHI}}},
\label{eq:magnet-reward}
\end{equation}

Table~\ref{tab:app:magnet} reports the results. Selecting among unsteered samples never satisfies
both conditions, because the unsteered model almost never reaches $\rho_{\mathrm{mag}} > 0.2$
(none of the 512 single samples). The CFG adapter gives the highest stable \& all, but its samples are
largely repeats of known magnets (37.9\% unique, 38.1\% novel), so few of them are SUN.
\ourmethod{} with $k{=}16$ matches the adapter on SUN \& all 
without training, while the adapter was fine-tuned on property-labelled data.

\begin{table}[h]
\caption{\textbf{Jointly targeting magnetization
density and HHI.} \emph{all} denotes the conjunction of the two on-target conditions. 
}
\label{tab:app:magnet}
\centering
\setlength{\tabcolsep}{2.5pt}
\resizebox{\linewidth}{!}{%
\begin{tabular}{lrrrrrrrrrr}
\toprule
 & & & & & & \multicolumn{3}{c}{on target} & & \\
\cmidrule(lr){7-9}
method & valid & unique & novel & stable & SUN & $\rho_{\mathrm{mag}}$ & HHI & all & stable \& all & SUN \& all \\
\midrule
MatterGen & 86.3 & 99.6 & 59.2 & 75.0 & \textbf{37.5} & 0.0 & 6.1 & 0.0 & $0.0${\scriptsize$\pm0.0$} & $0.0${\scriptsize$\pm0.0$} \\
$+$ CFG (joint adapter) & 80.3 & 37.9 & 38.1 & \textbf{79.9} & 14.1 & \Ba{49.2} & \textbf{63.9} & \textbf{34.0} & $\mathbf{31.1}${\scriptsize$\pm2.0$} & \Bam{5.5}{\scriptsize$\pm1.0$} \\
$+$ Best-of-N\keighta & \Bm{90.0} & \textbf{99.8} & \Bm{64.5} & 64.3 & 33.2 & 1.6 & 21.3 & 0.0 & $0.0${\scriptsize$\pm0.0$} & $0.0${\scriptsize$\pm0.0$} \\
$+$ Best-of-N ($k{=}16$) & 89.8 & \textbf{99.8} & 61.9 & 63.9 & 31.2 & 2.9 & 22.1 & 0.0 & $0.0${\scriptsize$\pm0.0$} & $0.0${\scriptsize$\pm0.0$} \\
$+$ \textbf{FK-ReMask}\keighta & 79.5 & 86.1 & 57.4 & 59.4 & 31.1 & 46.7 & 35.5 & 15.0 & $6.2${\scriptsize$\pm1.1$} & $3.1${\scriptsize$\pm0.8$} \\
$+$ \textbf{FK-ReMask} ($k{=}16$) & 76.2 & 74.2 & 59.4 & 47.9 & 25.0 & \textbf{63.7} & 43.2 & 30.3 & $11.1${\scriptsize$\pm1.4$} & $\mathbf{6.2}${\scriptsize$\pm1.1$} \\
\bottomrule
\end{tabular}
}
\end{table}

\end{document}

%% file: math_commands.tex
\usepackage{amsmath,amsfonts,bm}

\def\eqref#1{equation~\ref{#1}}

\def\1{\bm{1}}

\DeclareMathAlphabet{\mathsfit}{\encodingdefault}{\sfdefault}{m}{sl}
\SetMathAlphabet{\mathsfit}{bold}{\encodingdefault}{\sfdefault}{bx}{n}

\DeclareMathOperator*{\argmax}{arg\,max}